\documentclass[onecolumn,showpacs,showkeys,preprintnumbers,amsmath,amssymb]{revtex4}
\usepackage{graphicx}
\usepackage{dcolumn}
\usepackage{bm}
\usepackage{epsfig}
\usepackage{rotating}

\begin{document}
\title{Hierarchical Structure of the Foreign Trade: The Case of the United State}

\author{Ersin Kantar
\\
Department of Physics, Erciyes University, 38039 Kayseri, Turkey
}

\begin{abstract}
This study uses hierarchical structure methods (minimal spanning tree, (MST) and hierarchical tree, (HT)) to examine the hierarchical structures of the United State (US) foreign trade by using the real prices of their commodity export and import move together over time. We obtain the topological properties among the countries based on US foreign trade over the periods of 1985-2011. We also perform the bootstrap techniques to investigate a value of the statistical reliability to the links of the MSTs. Finally, we use a clustering linkage procedure in order to observe the cluster structure much better. The results of the topologies structural of these trees are as follows: i) We identified different clusters of countries according to their geographical location and economic growth. ii) Our results show that the European Union and Asian countries are more important within the network, due to a tighter connection with other countries. The country's most important trading partners are the Canada, China, Mexico, Japan, Germany, United Kingdom, South Korea, France, Taiwan, India, Singapore and Netherlands iii) We have also found that these countries play a significance role for US foreign trade and have important implications for the design of portfolio and investment strategies.
\end{abstract}

\keywords{Minimal Spanning Tree; Hierarchical Tree; Bootstrap Technique; Foreign Trade; United State}
\maketitle

\section{\label{sec:level1}Introduction}
The United States (US) is the most significant nation in the world when it comes to international trade. For decades, it has led the world in imports while simultaneously remaining as one of the top three exporters of the world. As the major epicenter of world trade, the United States enjoys leverage that many other nations do not. The US is a member of several international trade organizations. The purpose of joining these organizations is to come to agreement with other nations on trade issues, although there is domestic political controversy to whether or not the US government should be making these trade agreements in the first place. These organizations include: World Trade Organization, Organization of American States, and Security and Prosperity Partnership of North America. Main exports are: machinery and equipment, industrial supplies, non-auto consumer goods, motor vehicles and parts, aircraft and parts, food, feed and beverages. US imports non-auto consumer goods, fuels, production machinery and equipment, non-fuel industrial supplies, motor vehicles and parts, food, feed and beverages. Main trading partners are: Canada, European Union, Mexico, China and Japan.

The physicists have examined the statistical properties of stock market relationships (correlations, fluctuations,
etc.) in the price movements of different stocks to understand stock market dynamics. In this way, the
structure of stock markets has become one of the most intensively studied subjects in statistical physics and econophysics
\cite{Coelho2007a,Coelho2007b,Matteo2007,Podobnik2008,Yang2008,Wang2009a,Yamada2009,Racz2009,Plerou2009,Siven2009,Moro2009,Tabak2009a,Zhou2009,Qiu2009,Chakrabarti2009,Tabak2010a}. Moreover, real-life systems have been able to successfully describe the topological properties and characteristics \cite{Chen2006,Ray2007,Garas2008,Bivona2008,Wang2009b,Tabak2009b,Brida2010a,Kantar2011}. We present a study using the hierarchical structure methods, bootstrap technique, cluster linkage procedure on the basis of foreign trade data. In this paper we focus on foreign trade of the US and the main objective is to characterize the topology and taxonomy of the countries network. We should also mention that Podobnik et.al, studied annual logarithmic growth rates, R, of various economic variables such as exports and imports, and find that the distributions of R can be approximated by double exponential Laplace distributions in the central parts and power-law distributions in the tails \cite{Podobnik52008}. Moreover, the foreign trades ultimately diminished the differences between developed and developing countries; hence we suggested to state that some experts believe that convergence across all countries exists because of globalization \cite{Ausloos2007,Hidalgo2007,Podobnik62010}. Besides foreign trade many papers have reported a negative association between country wealth and corruption level, where on average richer countries are less corrupt \cite{Podobnik12007,Podobnik22008}. In addition to these, some papers reported that financial and economics time series of developed markets exhibit different scaling behavior than the series of undeveloped and developing markets \cite{Podobnik32006} Moreover, similar differences between developed markets and developing markets are reported also in financial series during large market crashes \cite{Podobnik42010}.

The aim of the present paper is to examine relationships
among countries based on US foreign trade by using the concept
of minimal spanning tree (MST) and hierarchical tree (HT) over the
period of 1985-2011. From these trees, both geometrical (through the
MST) and taxonomic (through the HT) information about the
correlation between the elements of the set can be obtained. Note
that the MST and then the HT are constructed using the Pearson
correlation coefficient as a measure of the distance between the
time series. Moreover, we performed bootstrap technique to associate
a value of reliability to the links of MSTs. We also used
the average linkage cluster analysis for obtaining the HT. These
methods give a useful guide to determining the underlying economic
or regional causal connections for individual countries.

The MST and HT introduced by Mantegna \cite{Mantegna1999}, and the structure of the currency markets \cite{Mizuno2006,Ortega2006,Naylor2007,Feng2007,Brida2009a,Feng2010,Keskin2010a}, world equity markets \cite{Bonanno2000,Coelho2007a}, European equity markets \cite{Gilmore2008} and commodity markets \cite{Sieczka2009,Tabak2010b} and time-varying behavior of stocks \cite{Onnela2003a,Onnela2003b,Onnela2002,Onnela2003c,Micciche2003,Coelho2007b} has been one of the intensively studied by using the Minimal Spanning Tree (MST) and the Hierarchical Tree (HT) tools. They have shown it to be useful in detecting clusters and taxonomic relations in a set of elements. We should mention that a bootstrap approach, which it has been used to
quantify the statistical reliability of hierarchical trees and correlation based networks Tumminello et al.
\cite{Tumminello2007a,Tumminello2007b,Tumminello2010}. In addition, Tumminello et al. \cite{Tumminello2011} investigated the statistical assessment of links in bipartite complex systems. Keskin \textit{et al.} \cite{Keskin2010a} studied the topology of correlation networks among 34 major currencies using the concept of a MST and HT for the full years of 2007-2008 when major economic turbulence occurred; they performed a technique to associate the value of statistical reliability to the links of the MSTs and HTs by using bootstrap replicas of data.  Moreover, Kantar \textit{et al.} applied the bootstrap technique to investigate the value of statistical reliability to the links on the hierarchical structures of Turkey's foreign trade \cite{Kantar2011} and the major International and Turkish companies \cite{Kantar2011a}. Finally, we should also mentioned that the average linkage cluster analysis, which observe the cluster structure much better in HTs \cite{Tumminello2005,Tumminello2007a,Kantar2011,Kantar2011a}.

The remainder of the paper is structured as follows. Next section introduces the methodology and the sampling procedures while Sec.
III shows the data and Sec. IV presents empirical results. Finally, Sec. V provides some final considerations.

\section{Methodology}
We determine input data that is needed to create the minimal spanning tree and the hierarchal tree. Then the test procedure in this paper can be explained by the following three steps. First, we create a countries network using the MST and the HT obtained starting from the MST. Second, bootstrap technique is used to establish statistical reliability and stability of our results. Finally, cluster structures are investigated by using a clustering linkage procedure.

\subsection{Minimal spanning tree and hierarchical trees}

Since the construction of a minimal spanning tree (MST) and hierarchical tree (HT), which is also called single linkage cluster analysis (SLCA) has been described extensively in Mantegna and Stanley \cite{Mantegna2000} as well as in our previous papers \cite{Keskin2010a,Kantar2011,Kantar2011a}, therefore we shall only give a brief summary here.

The correlation function between a pair of countries based on the debts of European countries in order to quantify synchronization between the countries is defined as

\begin{equation} \label{GrindEQ__1_}
C_{ij} =\frac{\left\langle R_{i} R_{j} \right\rangle -\left\langle R_{i} \right\rangle \left\langle R_{j} \right\rangle }{\sqrt{\left(\left\langle R_{i}^{2} \right. \rangle -\left\langle R_{i} \right\rangle ^{2} \right)\left(\left\langle R_{j}^{2} \right. \rangle -\left\langle R_{j} \right\rangle ^{2} \right)} } ,
\end{equation}
\noindent
where i and j are the numerical labels to the debts of countries and the notation $\left\langle ...\right\rangle $ means an average over time. ${ R}_{{ i}} (t)$ is defined as ${ R}_{{ i}} {(t)\; =\; ln\; P}_{{ i}} { (t\; +\; }\tau { )\; -\; ln\; P}_{{ i}} { (t)}$ where ${ P}_{{ i}} (t)$ is the rate i at the time $\tau$. All cross-correlations range from -1 to 1, where -1 and +1 mean that two countries i and j are completely anti-correlated and correlated, respectively. In the case of $C_{ij} $ = 0 the countries i and j are uncorrelated.

The MST is based on the idea that the correlation coefficient between a pair of countries can be transformed to a distance between them by using an appropriate function as a metric. An appropriate function for this transformation is

\begin{equation} \label{GrindEQ__2_}
{\rm d}_{{\rm ij}} =\sqrt{2(1-C_{ij} )} ,
\end{equation}
\noindent
where ${\rm d}_{{\rm ij}} $ is a distance for a pair of the rate i and the rate \textit{j}. Now, one can construct an MST for a pair of countries using the N x N matrix of ${\rm d}_{{\rm ij}} $.

On the other hand, to construct an HT, we introduce the ultrametric distance or the maximal ${\rm d}_{{\rm ij}}^{{\rm \wedge }} $${}_{ }$ between two successive countries encountered when moving from the first country \textit{i} to the last country \textit{j} over the shortest part of the MST connecting the two countries. The distance fulfills the condition ${\rm d}_{{\rm ij}}^{{\rm \wedge }} {\rm \; }\le {\rm max\{ d}_{{\rm ik}} {\rm ,\; d}_{{\rm kj}} {\rm \} }$, which is a stronger condition than the usual triangular inequality ${\rm d}_{{\rm ij}}^{{\rm \wedge }} {\rm \; }\le {\rm \; d}_{ik}^{{\rm \wedge }} {\rm \; +\; d}_{{\rm kj}}^{{\rm \wedge }} $ \cite{Rammal1986}. The distance ${\rm d}_{{\rm ij}}^{{\rm \wedge }} $ is called the subdominant ultrametric distance \cite{Benzecri1984}. Then, one can construct an HT by using this inequality.

We also use average linkage cluster analysis (ALCA) in order to observe the different clusters of countries according to their geographical location and economic growth more clearly. Since the procedures to obtain ALCA have been presented by Tumminello et al. \cite{Tumminello2010} and Kantar et al. \cite{Kantar2011,Kantar2011a} in detail, we will not explain its construction in here.

\subsection{The measure of link reliability with bootstrap technique}

In correlation based hierarchical investigations, the statistical reliability of hierarchical trees and networks depends on the statistical reliability of the sample correlation matrix. Laloux et al. \cite{Laloux1999} and Plerou et al. \cite{Plerou1999} applied Random Matrix Theory methods to obtain the quantitative estimation of the statistical uncertainty of the correlation matrix. On the other hand, the bootstrap technique, which was invented by Efron \cite{Efron1979}, has been widely used in phylogenetic analysis since the paper by Felsenstein \cite{Felsenstein1985} as a phylogenetic hierarchical tree evaluation method \cite{Efron1996}. This technique was used to quantify the statistical reliability of hierarchical trees and correlation based networks by Tumminello et al. \cite{Tumminello2007b,Tumminello2007a,Tumminello2010}. Kantar et al. also applied the bootstrap technique to investigate the value of statistical reliability to the links on the hierarchical structures of Turkey's foreign trade \cite{Kantar2011} and major international and Turkish companies \cite{Kantar2011a}.

In order to quantify the statistical reliability of the links of the MST the bootstrap technique is applied to the data. The numbers appearing in Fig. 1 quantify this reliability (bootstrap value) and they represent the fraction of replicas preserving each link in the MST. Developed by Efron in 1979, the bootstrap method has been recently proposed as a technique to measure the reliability of the links of minimum spanning trees and hierarchical trees obtained with financial data \cite{Tumminello2007a,Tumminello2007b}. We should also mention that this technique has been well explained in Refs. \cite{Keskin2010a,Kantar2011,Tumminello2007a,Tumminello2007b,Tumminello2010}.

\section{Data}

We chosen 66 countries from the different continents in US foreign trade. We used the data period from January 1st 1985 to December 30 th 2011 and listed countries, continent and their corresponding symbols given in Table 1. The monthly prices downloaded from "The United States Census Bureau" database, available online (http://www.census.gov). We consider all commodities, i.e.,  the machinery and mechanical appliances, vehicle for railways, electrical machinery and equipment: parts thereof, iron and steel, mineral fuels, minerals oils and product of their distillation, etc. Foreign trade measured by the US Dollar. We will construct the hierarchical trees from these data in the next section.

\section{Numerical Results and Discussions}

In this section, we present the MSTs, including the bootstrap values, and the HTs of 66 countries based on the US foreign trade
from 1985-2011 period. We also investigate cluster structures by using a clustering linkage procedure. We
choose the US as numeraire which is one of most significant nation in the world when it comes to international trade.
We construct the MSTs by using Kruskal's algorithm \cite{West1996,Kruskal1956,Cormen1990} for the US foreign
trade based on a distance-metric matrix. The amounts of the links that persist from one node(country) to the other correspond to the relationship between the countries of Turkey's foreign trade. We carried out bootstrap technique to associate a value of the statistical reliability to the
links of the MSTs. If the values are close to one, the statistical reliability or the strength of the link is very high.
Otherwise, the statistical reliability or the strength of the link is lower \cite{Keskin2010a, Tumminello2007a}. We also obtained
cluster structure of the hierarchical trees much better by using average linkage cluster analysis.

\subsection{Examination of the structure of exports in the United State}

\subsubsection{The MST}

Fig. 1(a) shows the MST applying the method of Mantegna \cite{Mantegna1999}, Mantegna and Stanley \cite{Mantegna2000} for
the export based on a distance-metric matrix for the period 1985-2011. In Fig. 1(a), we observed different clusters of countries according to their geographical proximity and economic growth. In this figure, we detected three different cluster, namely the European, Asian and America countries. It can also be clearly seen that in the MST, the European Union countries form the central structure. It is observed that DEU is in the center of European Union countries and it is prodominant countries for this period. First cluster consist of DEU, FRA, ITA, DNK, PRT, NLD, ESP, BEL, IRL, FIN, GBR, SWE and NOR. In this cluster, there are a strong relationship among DEU - FRA, DEU - BEL, FRA - DNK, DEU - ISL and FRA - GBR. We can establish this fact from the bootstrap value of the link between these countries, which are equal to 1.00, 0.89, 0.87, 0.86 and 0.83 in a scale from zero to one, respectively; hence these countries are very closely connected with each other. The second cluster formed the majority of Asian countries, and separated two sub-groups. The first sub-group contains SGP, KOR, TWN, EGY and IDN, there is a strong relationship between the SGP and KOR. We can find this fact from the bootstrap value of the link between the SGP and KOR, which is equal to 0.89 in a scale from zero to one. The second sub-group is consist of the JPN, CHN and MYS. On the other hand, the bootstrap values of the links between the JPN - CHN and SGP - EGY are very low. This means that these links could only a statistical fluctuation. The third cluster are composed of the America countries, namely, CAN, MEX, DOM and JAM. In this cluster, there are a strong relationship among the CAN - MEX and DOM - JAM. We can establish this from the bootstrap value of the links among the countries, which are equal to 0.79 and 0.78 in a scale from zero to one, respectively.

\subsubsection{The HT}

The HT of the subdominant ultrametric space associated with the MST is shown in Fig. 1(b). Two countries (lines) link when a horizontal
line is drawn between two vertical lines. The height of the horizontal line indicates the ultrametric distance at which the two
countries are joined. To begin with, in Fig. 1(b), we can observe three clusters. First cluster is composed to European countries cluster, which is consists of DEU, GBR, CHE, ESP, DNK, BEL, AUT, FRA, SWE, BEL, NLD and PRT. The distance between the SWE and BEL is the smallest of the sample, indicating the strongest relationship between these two countries. The second cluster is the Asian countries cluster. In this cluster, the distance between TWN and KOR is the smallest of the sample, indicating the strongest relationship between these two countries. The third cluster is composed of the American countries, which is consists of MEX and CAN.

\subsubsection{The ALCA}

In HT, we used the average linkage cluster analysis (ALCA) in order to observe the cluster structure much better. The HT seen in Figs. 1(c) is obtained for the based on foreign trade for the period 1985-2011. When comparing the HT and ALCA, similar cluster structures were observed; however, the number of countries and clusters in the ALCA cluster was found to be more than in the HT. For example, thirteen countries in the cluster  of composed to Asian were seen in the HT, but fourteen countries were seen in ALCA, as can be verified comparing Fig. 1(b) with Fig. 1(c).
\subsection{Examination of the structure of imports in the United State}

\subsubsection{The MST}

Fig. 2(a) shows the MST for the imports based on a distance-metric matrix for the period 1985-2011. We see that the cluster structures are obtained more efficiently in obtaining MST for imports. In this figure, we detected three different cluster, namely the European, Asian and America countries. It can also be clearly seen that in the MST, the European and Asian countries form the central structure. First cluster consist of DEU, FRA, AUT, ITA, DNK, PRT, NLD, CHE, ESP, BEL, IRL, GBR and SWE. In this cluster, there are a strong relationship among DEU - ESP, SWE - BEL, IRL - DNK, DEU - FRA and DEU - NLD. We can establish this fact from the bootstrap value of the link between these countries, which are equal to 0.98, 0.97, 0.95, 0.92 and 0.87 in a scale from zero to one, respectively; hence these countries are very closely connected with each other. The bootstrap values of the links between the FIN - KOR and POL - THA are very low, which could only a statistical fluctuation. Second cluster consist all Asian countries in Table 1. In this cluster, there are a strong relationship among TWA - KOR, TWA - HKG, TWA - THA and HKG - MAC. We can establish this fact from the bootstrap value of the link between these countries, which are equal to 1.00; hence these countries are very closely connected with each other. Moreover, the bootstrap values of the links between the TWA - EGY, JPN - THA and IDN - KWT are very low. This means that these links could only a statistical fluctuation. The third cluster formed the majority of American countries, and separated two sub-groups. The first sub-group contains DOM, HTI, CRI, HND, SLV and JAM, there is a strong relationship between the DOM and HTI. We can find this fact from the bootstrap value of the link between the DOM and HTI, which is equal to 0.95. The second sub-group is consist of the MEX, CAN and PRY.

\subsubsection{The HT}

The HT of the subdominant ultrametric space associated with the MST is shown in Fig. 2(b). Two countries (lines) link when a horizontal
line is drawn between two vertical lines. The height of the horizontal line indicates the ultrametric distance at which the two
countries are joined. To begin with, in Fig. 2(b), we can observe three clusters. First cluster is composed to American countries cluster, which is consists of the two sub-group, namely first group (MEX and CAN) and  second group (HND and SLV). The second cluster is the European countries. The  third cluster is composed of the Asian countries, which is consists of SGP, TWN and KOR.

\subsubsection{The ALCA}

In HT, we used the average linkage cluster analysis (ALCA) in order to observe the cluster structure much better. The HT seen in Figs. 2(c) is obtained for the based on foreign trade for the period 1985-2011. When comparing the HT and ALCA, similar cluster structures were observed; however, the number of countries and clusters in the ALCA cluster was found to be more than in the HT. For example, four countries in the cluster of composed to European were seen in the HT, but six countries were seen in ALCA, as can be verified comparing Fig. 2(b) with Fig. 2(c).

Finally, it is worth mentioning that according to the United States Census Bureau, the fifteen largest trading partners of the United States represent 75.0\% of U.S. imports, and 72.5\% of U.S. exports as of December 2010. The United State's most important trading partners are the CAN, CHN, MEX, JPN, DEU, GBR, KOR, FRA, TWN, NLD, SGP, IND and VEN. Moreover, as a single economy, the EU is the largest trading partner of the US with \$319.6 billion worth of EU goods going to the US and \$239.8 billion of US goods going to the EU as of 2010, totaling approximately \$559.4 billion in total trade.

\section{SUMMARY AND CONCLUSION}
We presented the hierarchical structures of countries based on
foreign trade of United State by using the concept of the MST, including the bootstrap values, and the HT
for the 1985-2011 periods. We obtained the clustered structures of the trees and identified
different clusters of countries according to their geographical location and economic ties. From the topological structure of these
trees, we found that the European and Asian countries United State's most important trading partners, specifically DEU, GBR, FRA, NLD, JPN, KOR, TWN, SGP and IND are located in the center of the MSTs. We performed the bootstrap technique to associate a value of statistical reliability to the links of MSTs for getting the information about statistical reliability of each link of trees. From the results of
the bootstrap technique, we can see that in general, the bootstrap
values in MSTs are highly consistent with each other. We
also used the average linkage cluster analysis for obtaining cluster
structure much better of the hierarchical trees. Finally, we expect
that the present paper helps for a better understanding of overall
structure of US foreign trade, and also provide a valuable
platform for the theoretical modeling and further analysis. The same analysis will also successfully apply to the other developed markets; hence in a future study, we aim to perform the analysis to investigate the developed markets.

\begin{center}
\textbf{List of the Figure Captions}
\end{center}

\begin{itemize}
\item [\textbf{Fig. 1(a)}] Minimum spanning tree associated to monthly data of the 66 countries in United State export during the 1985-2011 period.
\item [\textbf{(b)}] The hierarchical tree corresponds to the MST in the years of the 66 countries in United State export during the 1985-2011 period.
\item [\textbf{(c)}] Average linkage cluster analysis. Hierarchical tree associated to a system of the 66 countries in United State export during the 1985-2011 period.

\item [\textbf{Fig. 2(a)}] Minimum spanning tree associated to monthly data of the 66 countries in United State import during the 1985-2011 period.
\item [\textbf{(b)}] The hierarchical tree corresponds to the MST in the years of the 66 countries in United State import during the 1985-2011 period.
\item [\textbf{(c)}] Average linkage cluster analysis. Hierarchical tree associated to a system of the 66 countries in United State import during the 1985-2011 period.

\end{itemize}

\begin{center}
\textbf{List of the Table Captions}
\end{center}

\begin{itemize}
\item [\textbf{Table 1}] Countries, continents and their corresponding symbols.
\end{itemize}
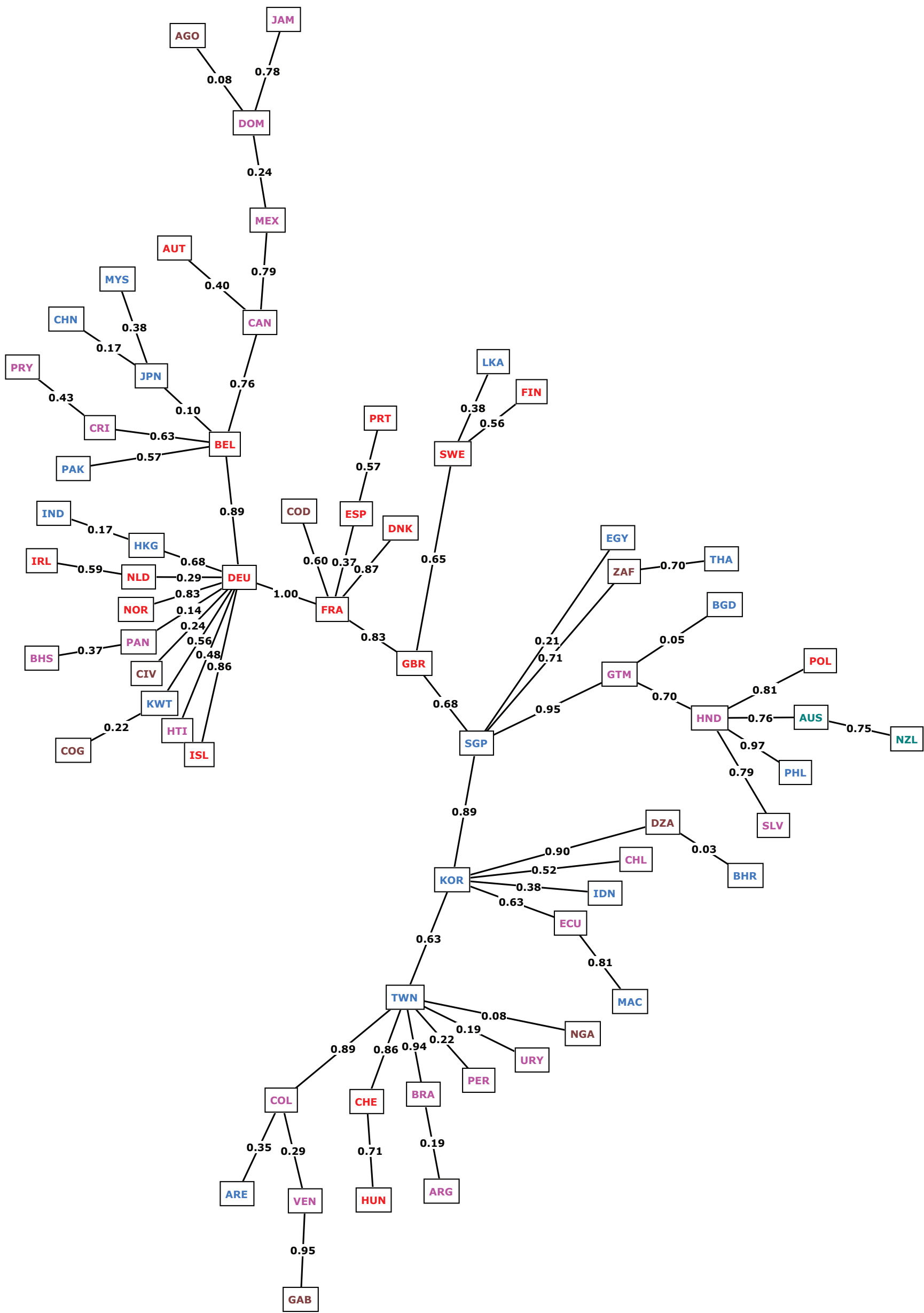
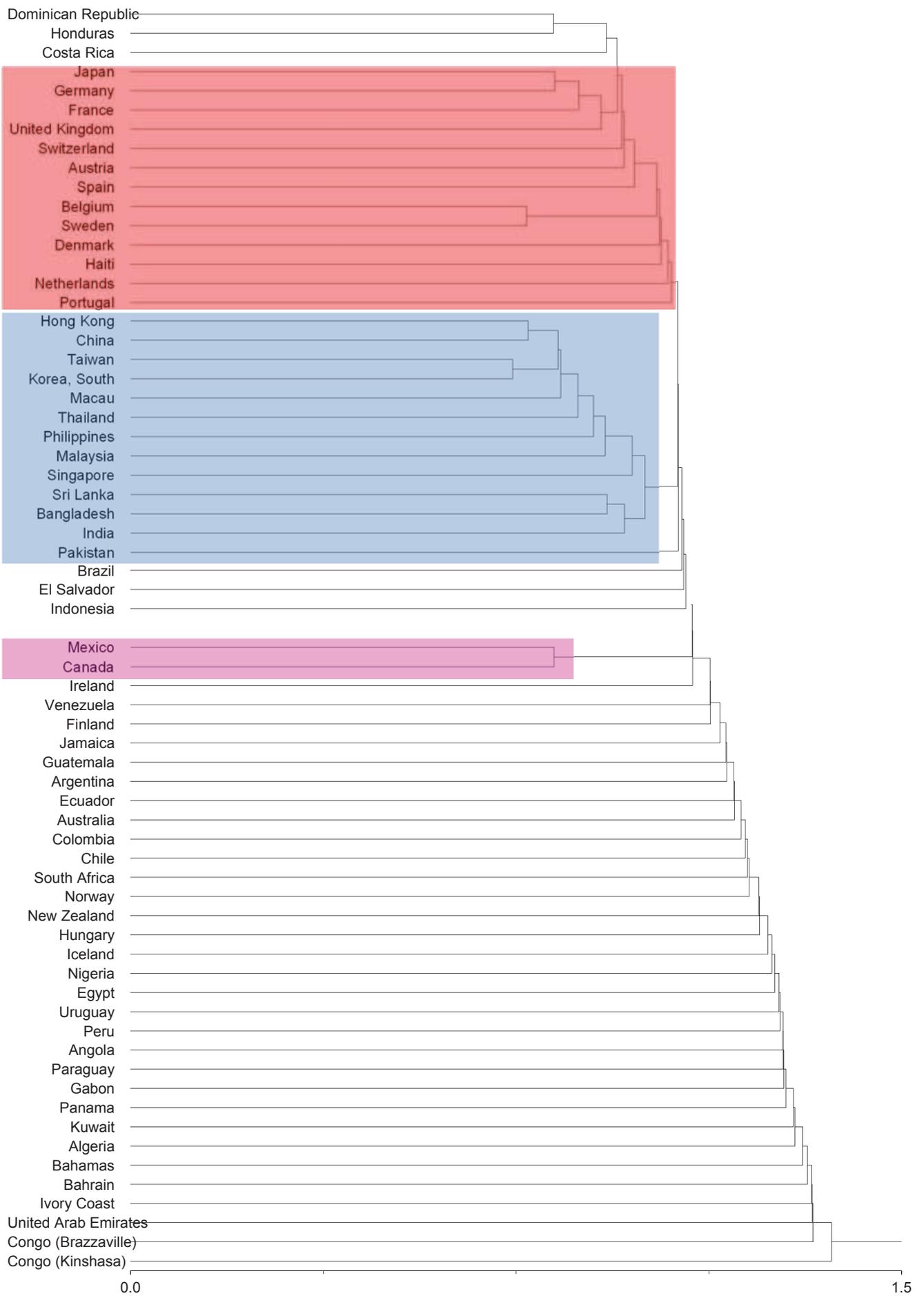
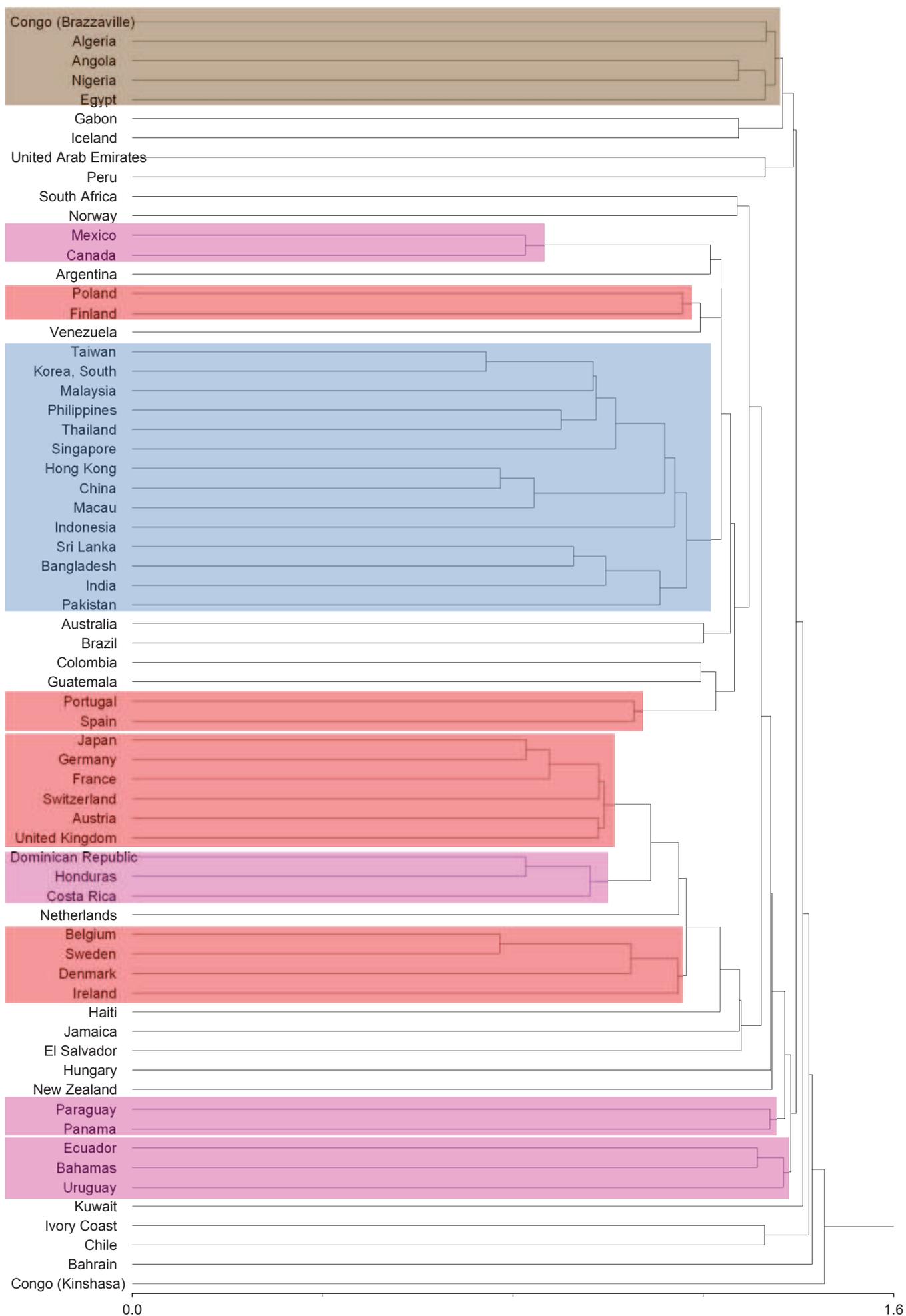
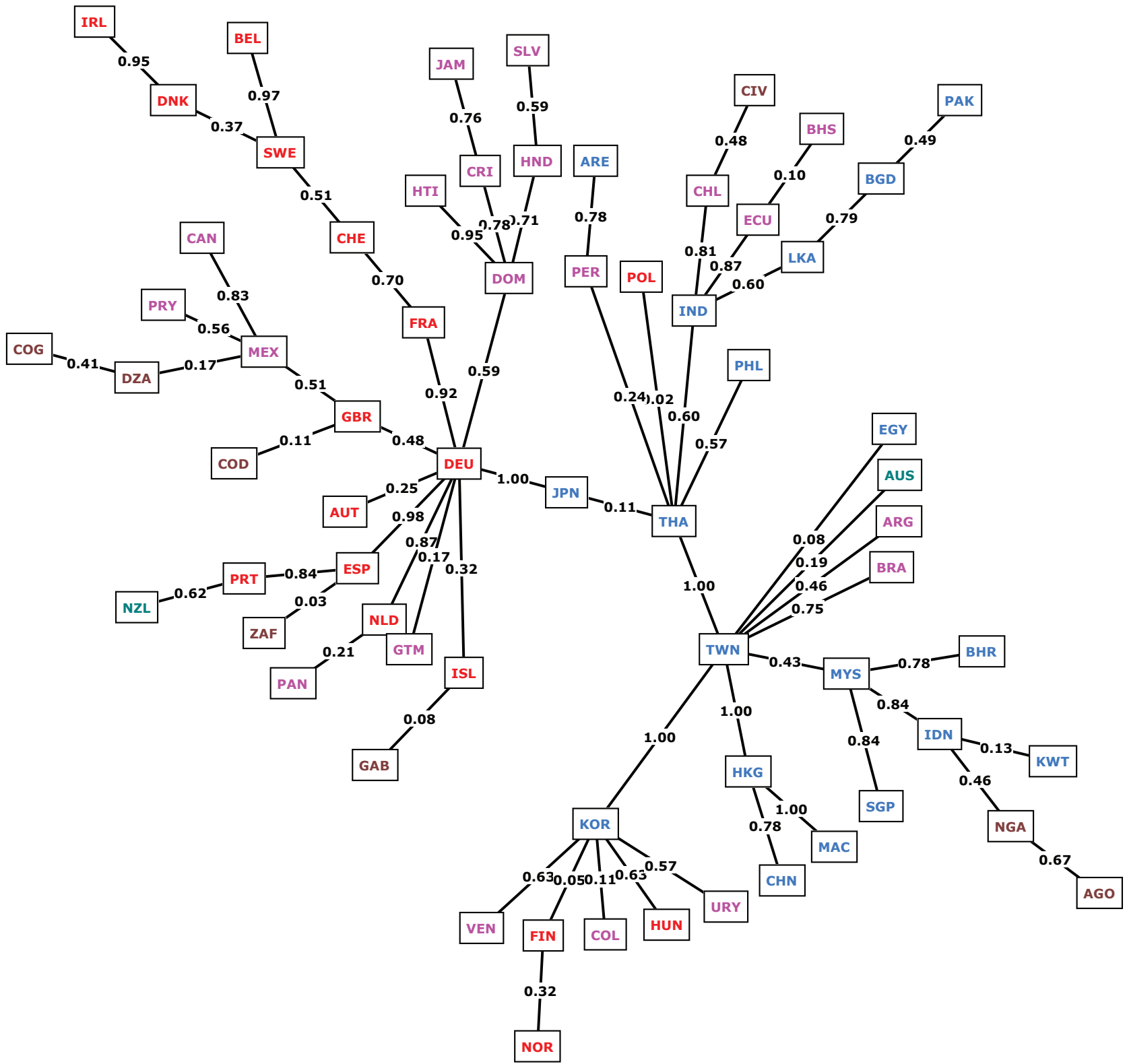
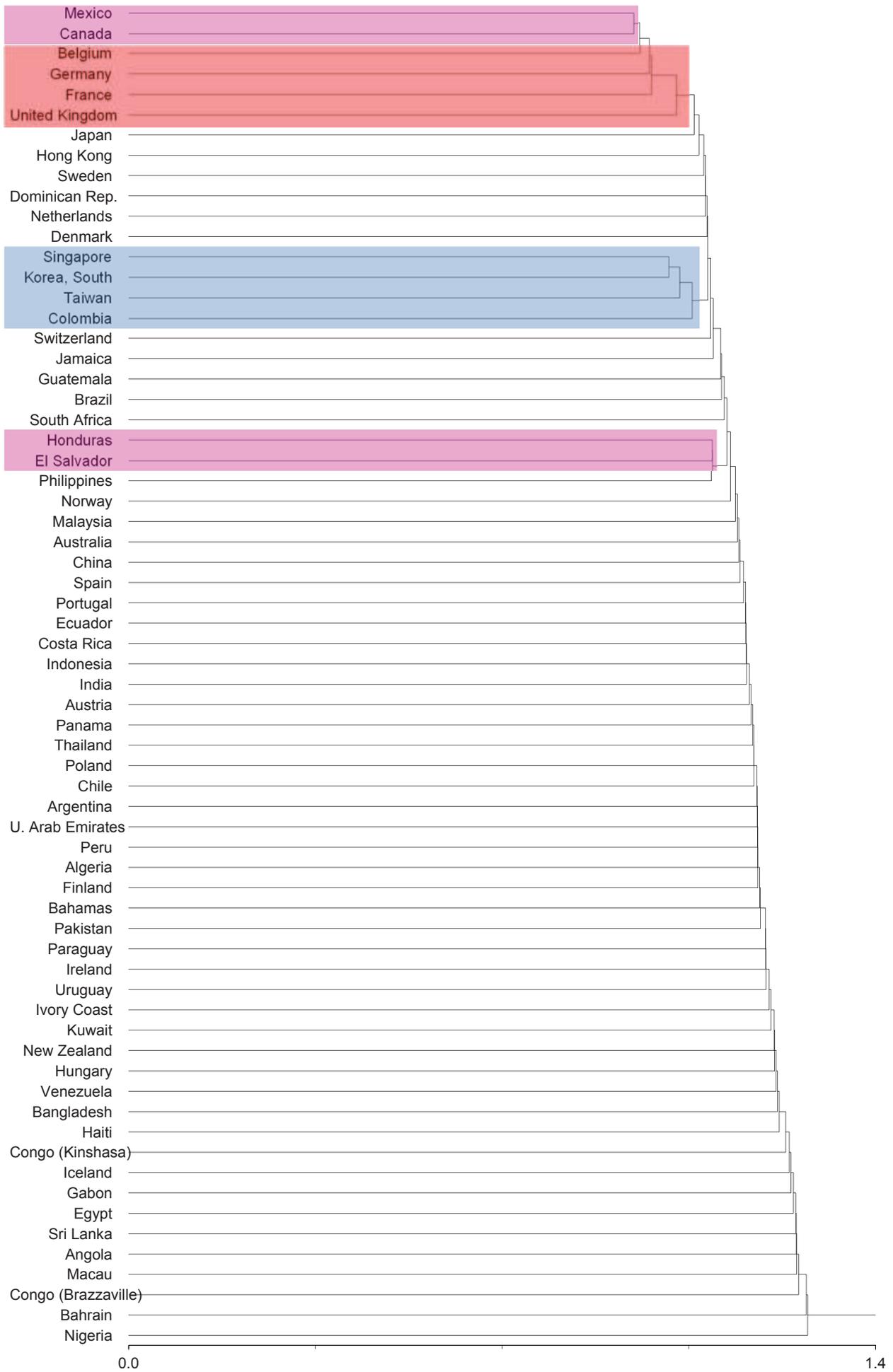
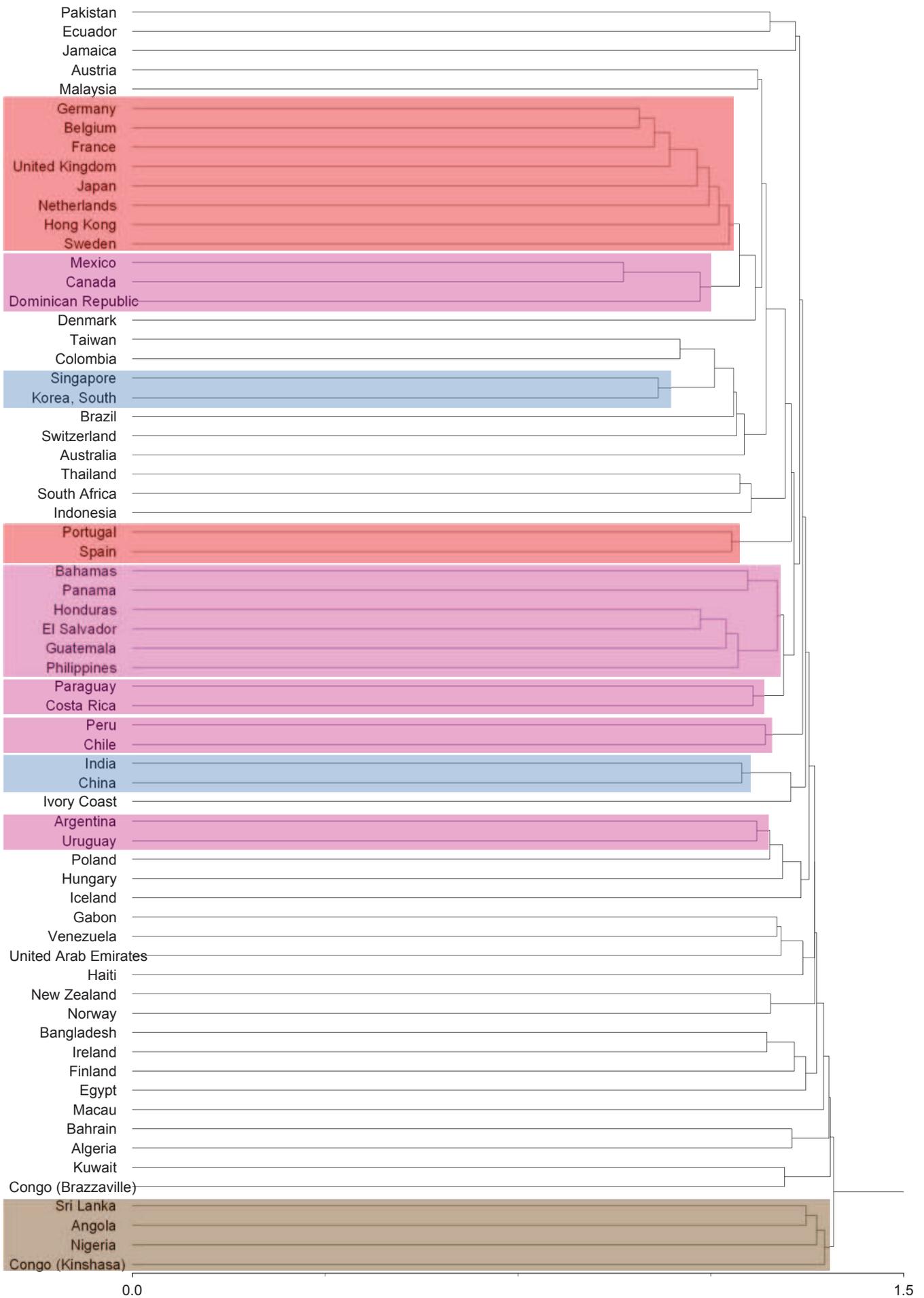
\end{document}